\documentclass[english,aps,manuscript]{revtex4}
\usepackage[T1]{fontenc}
\usepackage[latin9]{inputenc}
\usepackage{bm}
\usepackage{amsmath}
\usepackage{amssymb}
\usepackage{graphicx}
\usepackage{color}

\makeatletter
\@ifundefined{textcolor}{}
{%
 \definecolor{BLACK}{gray}{0}
 \definecolor{WHITE}{gray}{1}
 \definecolor{black}{rgb}{1,0,0}
 \definecolor{GREEN}{rgb}{0,1,0}
 \definecolor{BLUE}{rgb}{0,0,1}
 \definecolor{CYAN}{cmyk}{1,0,0,0}
 \definecolor{MAGENTA}{cmyk}{0,1,0,0}
 \definecolor{YELLOW}{cmyk}{0,0,1,0}
}

\usepackage{babel}

\makeatother

\usepackage{babel}
\begin{document}
\title{Drag Coefficient in Near-Critical Binary Mixtures: Solving Hydrodynamic Fields with Improved Numerics}
\author{Shunsuke Yabunaka}
\email{yabunaka123@gmail.com}
\affiliation{Advanced Science Research Center, Japan Atomic Energy Agency, Tokai,
319-1195, Japan}
\date{\today}
\begin{abstract}
We calculate the drag coefficient of a spherical particle suspended
in a near-critical binary fluid mixture. To capture the scaling behavior
associated with critical adsorption in the strong adsorption regime,
we employ the framework of local renormalized functional theory. Previous
theoretical studies encountered numerical difficulties when attempting
to solve the coupled hydrodynamic and chemical potential equations,
expressed as integral equations, for systems with large bulk correlation
lengths. These difficulties limited direct comparison with experimental
results. In this study, we overcome those limitations by reformulating
the hydrodynamic equations as a set of ordinary differential equations
using a compactified radial coordinate. This approach enables more
stable numerical computation and facilitates the implementation of
appropriate boundary conditions at large distances from the particle.
As a result, we successfully compute the drag coefficient over a broader
range of bulk correlation lengths than in previous works and compare our
theoretical predictions with available experimental data. 
\end{abstract} 
\maketitle

\section{Introduction}

It is well known from Stokes' law \citep{stokes1851effect} that a
spherical particle of radius $a$ experiences a drag force $F=\bar{\gamma}v$
when it moves slowly with velocity $v$ in an incompressible, one-component
fluid of shear viscosity $\bar{\eta}$. The drag coefficient is given
by $\bar{\gamma}=6\pi\bar{\eta}a$. According to Einstein, the diffusion
constant $D$ of such a particle is related to the drag coefficient
through the Stokes-Einstein relation: $D$ equals {}{$k_{B}T/\bar{\gamma}$}
where $k_{B}$ is the Boltzmann constant \citep{einstein1905motion}.

Deviations from Stokes' law can arise in complex fluids or soft materials,
where the surrounding medium exhibits structural or dynamic heterogeneities.
In microrheology, a colloidal particle is employed as a probe to infer
local rheological properties at mesoscopic scales \citep{zia2018active}.
Several theoretical studies have investigated such deviations in specific
systems, such as in isotropic phases of nematic liquid crystals with
anchoring effects \citep{fukuda2005friction} and in semidilute polymer
solutions \citep{furukawa2004microrheology}.

In this work, we focus on the drag force acting on a spherical particle
immersed in a near-critical binary fluid mixture. {}{T}he particle surface generally exhibits preferential adsorption
of one component, resulting in the formation of an adsorption layer.
{}{Near the critical
point, this} effect becomes particularly pronounced close to the critical temperature
\citep{beysens1985adsorption}.  This adsorption layer not only plays
a crucial role in particle interactions within colloidal suspensions  \cite{furukawa2013nonequilibrium, okamoto2013attractive}, which should induce reversible aggregation of colloidal particles observed in experiments \cite{beysens1985adsorption, gallagher1992partitioning, PhysRevLett.100.188303, bonn2009direct},  {}{but also in} osmotic effects of a near critical binary mixture inside a capillary tube \cite{wolynes1976osmotic, PhysRevE.109.064610, samin2017interplay}. When the particle moves, this adsorption
layer is deformed, and the resulting inhomogeneity in the surrounding
fluid can exert an additional force on the particle beyond the conventional
viscous drag.

Experimentally, it has been observed that the drag coefficient $\bar{\gamma}$
increases almost linearly with the bulk correlation length $\xi_{\infty}$
{}{typically} for $0.3\lesssim\xi_{\infty}/a\lesssim2$, as estimated from the Stokes-Einstein
relation \citep{omari2009effect}. This has been interpreted as an
effective increase in the hydrodynamic size of the particle due to
the adsorption layer. However, this interpretation is not fully justified,
as the adsorption layer cannot be treated as a rigid object.

From a theoretical perspective, the strong adsorption regime near
criticality requires the inclusion of renormalization effects. Renormalization
group studies have revealed universal profiles of critical adsorption
layer \citep{floter1995universal, hanke1999critical}, showing that,
inside the layer, the composition deviation decays algebraically from
the surface, with a typical thickness of the order of $\xi_{\infty}$
\citep{rudnick1982order}. To capture this behavior within a free
energy framework, Fisher and de Gennes proposed a local renormalized
functional theory at the critical temperature $T=T_{c}$ \citep{fisher1978wall},
which was later extended to $T\neq T_{c}$ by Okamoto and Onuki \citep{okamoto2012casimir}. {}{To calculate the enthalpy density and describe off-critical regions very close to the wall more precisely, the non-random two liquid model is employed as well \citep{FUJITANI2024114050}.}

In earlier work, Okamoto et al. computed the drag coefficient using
a Gaussian-type free energy density, valid for weak adsorption, and
found that the deviation $\Delta\bar{\gamma}$ scales as $\xi_{\infty}^{6}$
for $\xi_{\infty}/a\lesssim1$ \citep{okamoto2013drag}. This prediction
is significantly steeper than the linear trend observed experimentally.
More recently, calculations based on the renormalized functional theory
have reproduced a much more gradual, nearly linear dependence for
$0.3\lesssim\xi_{\infty}/a\lesssim0.6$ \citep{yabunaka2020drag}.
However, for larger bulk correlation lengths, previous studies encountered
numerical difficulties when solving the integral equations equivalent
to the hydrodynamic equations, preventing accurate comparison with
experiments.

In this paper, we directly solve the hydrodynamic equations as a set
of ordinary differential equations (ODEs) using a compactified radial
variable, similar to that introduced in \citep{fukuda2005friction}.
This formulation facilitates the imposition of appropriate asymptotic
behaviors at infinity and enables us to compute the drag coefficient
over a wider range of $\xi_{\infty}$. Our results are compared with
experimental data to provide improved insight into the hydrodynamic
response in near-critical mixtures. 

\section{Formulation}

We briefly summarize the formulation to calculate the drag coefficient
of a spherical particle in a near critical mixture with local renormalized
functional theory. See for \citep{okamoto2013drag,yabunaka2020drag}
the detailed derivations. The Boltzmann constant $k_{B}$ will be
set to unity.

\subsection{Statics}

The order parameter $\psi$ is {}{defined as} $\varphi - \varphi_c$, where $\varphi$ is the local composition and $\varphi_c$ is the critical composition. The reduced temperature $\tau$ is defined as $\left(T - T_c\right)/T_c$. In our calculation, we set the critical exponents to the following values {}{\citep{onuki2002phase}}
\begin{equation}
\alpha = 0.110, \; \beta = 0.325, \; \gamma = 1.240, \; \nu = 0.630, \; \eta = 0.0317.
\end{equation}
We denote the fluid region outside the particle as $C^{e}$ and the surface of the particle as $\partial C$. In the presence of the solid surface of the particle, the free energy functional consists of bulk and surface parts as
\begin{equation}
\mathcal{F} = \int_{C^{e}} d\boldsymbol{r} f + \int_{\partial C} dS f_s.
\end{equation}
The bulk free energy density is
\begin{equation}
f = T_c \left[ \frac{r}{2}\psi^{2} + \frac{u}{4}\psi^{4} + C\left|\nabla\psi\right|^{2} \right],
\end{equation}
where the singular contribution for $\psi = 0$ proportional to $\left|\tau\right|^{2-\alpha}$, yielding the specific heat singularity, is neglected. The chemical potential field is defined as
\begin{equation}
\mu\left(\boldsymbol{r}\right) = \frac{\delta \mathcal{F}}{\delta \psi\left(\boldsymbol{r}\right)}.
\end{equation}
In the local renormalized functional theory, in terms of a nonnegative parameter $w$ representing the distance from the critical point in the $\tau$-$\psi$ plane, $r$, $u$, and $C$ are given by
\begin{align}
\frac{r}{\tau} & = C_{1}\xi_{0}^{-2} w^{\gamma-1}, \\
\frac{u}{u^{*}} & = C_{1}^{2}\xi_{0}^{-1} w^{\left(1-2\eta\right)\nu}, \\
C & = C_{1} w^{-\eta\nu},
\end{align}
where $C_{1}$ and $u^{*}$ are constants. We consider the one-phase region and set $\tau>0$. $u^{*}$ is a universal constant, and we set $u^{*} = 2\pi^{2}/9$, obtained by the 1-loop calculation. $w$ is locally determined as a function of $\tau$ and $\psi$ by
\begin{equation}
w = \tau + 3u^{*}C_{1}\xi_{0} w^{1-2\beta}\psi^{2}.
\end{equation}
The correlation length for uniform $\psi$ becomes $\xi = \xi_{0} w^{-\nu}$, and we define the local correlation length $\xi = \xi_{0} w^{-\nu}$ also for nonuniform profiles of $\psi$. We assume the surface free energy density as
\begin{equation}
\frac{f_{s}}{T_{c}} = -h\psi,
\end{equation}
where $h$ is the surface field. Here we neglect the surface critical exponents $\Delta_{1}$ and $\beta_{1}$ associated with the critical fluctuation very close to the surface, which play an important role for small $h$.

We assume that the composition field approaches the critical composition
\begin{equation}
\psi\left(\boldsymbol{r}\right) \rightarrow 0 \; \text{as}\; \left|\boldsymbol{r}\right| \rightarrow \infty. \label{eq:boundary_r_infty_psi}
\end{equation}
The equilibrium profile $\psi^{\left(0\right)}\left(\boldsymbol{r}\right)$ can be obtained by minimizing the free energy functional. The equilibrium conditions are
\begin{equation}
\mu\left(\boldsymbol{r}\right) = 0 \; \text{(in fluid)},
\end{equation}
\begin{equation}
C\boldsymbol{n}\cdot\nabla\psi = h \; \text{(on surface)}, \label{eq:surface_bc}
\end{equation}
where $\boldsymbol{n}$ is the unit normal vector on the solid surface $\partial C$, pointing outside of the particle. We assume that Eq.~\eqref{eq:surface_bc} holds also in dynamics, as done in previous studies {}{\citep{puri1997surface}}.

\subsection{Dynamics}

{}{Here, we introduce a dimensionless smallness
parameter $\varepsilon$ and expand the dynamic equations up to the
first order in $\varepsilon$.} We consider the steady state where an external force $\varepsilon E\bm{e}_{z}$
is acting on a spherical particle of radius $a$ immersed in a quiescent
near-critical mixture whose composition approaches the critical one
far from the particle. The particle moves at a constant velocity $\varepsilon U$
in the $z$-direction.  The Reynolds number and $\varepsilon U\xi_{\infty}(L(\psi=0)f''(\psi=0))^{-1}$
must be small to justify the linear order calculation.

We denote the velocity field by $\bm{v}$ and assume that the surrounding
fluid is incompressible:

\begin{equation}
\nabla\cdot\boldsymbol{v}=0.
\end{equation}
The boundary conditions for the velocity field are 
\begin{equation}
\boldsymbol{v}\left(\boldsymbol{r}\right)\rightarrow0\:\left(\left|\boldsymbol{r}\right|\rightarrow\infty\right),
\end{equation}
\begin{equation}
\boldsymbol{v}\left(\boldsymbol{r}\right)=\varepsilon U\boldsymbol{e}_{z}\;\mathrm{on}\:\partial C,\label{eq:no-slip}
\end{equation}
where the non-slip condition is assumed on the surface of the particle.
The composition field $\psi$ satisfies 
\begin{equation}
\frac{\partial\psi}{\partial t}=\boldsymbol{v}\cdot\nabla\psi+\nabla\cdot\left[L\left(\psi\right)\nabla\mu\right].\label{eq:dpdt}
\end{equation}
The kinetic coefficient $L\left(\psi\right)$ is set to 
\begin{equation}
L\left(\psi\right)=\frac{T}{6\pi\bar{\eta}\xi f''\left(\psi\right)},
\end{equation}
which depend on $\tau$ and $\psi$ and reproduces the expression
of diffusion constant 
\begin{equation}
D_{\xi}=\frac{T}{6\pi\bar{\eta}\xi},
\end{equation}
predicted by Kawasaki when the composition is the critical one $\psi=0$
everywhere in space and $\tau>0$ {}{\citep{KAWASAKI19701}}. Here $\bar{\eta}$ represents the
viscosity, which we assume to be constant neglecting its composition
dependence and weak divergence near the critical point. The boundary
conditions for the chemical potential field are 
\begin{equation}
\mu\left(\boldsymbol{r}\right)\rightarrow0\:\left(\left|\boldsymbol{r}\right|\rightarrow\infty\right),
\end{equation}
\begin{equation}
\boldsymbol{n}\cdot\nabla\mu=0\;\mathrm{on}\:\partial C,\label{eq:nmupr}
\end{equation}
where the former one guarantees Eq. (\ref{eq:boundary_r_infty_psi})
and the latter one holds since there is no {}{diffusive} composition flux penetrating
the surface $\partial C$. The velocity field obeys the following
Stokes equation 
\begin{equation}
0=\bar{\eta}\nabla^{2}\boldsymbol{v}-\psi\nabla\mu-\nabla p,\label{eq:stokes eq-1}
\end{equation}
where $p$ is the pressure field determined from the incompressibility
condition. In our calculation, we neglect the noise terms in the hydrodynamic
equations in Eqs. (\ref{eq:dpdt}) and (\ref{eq:stokes eq-1}), assuming
the renormalization effects are effectively taken into account the
static free energy functional and the dynamic coefficients $L\left(\psi\right)$
and $\bar{\eta}$.

\begin{figure}
\includegraphics[scale=0.5]{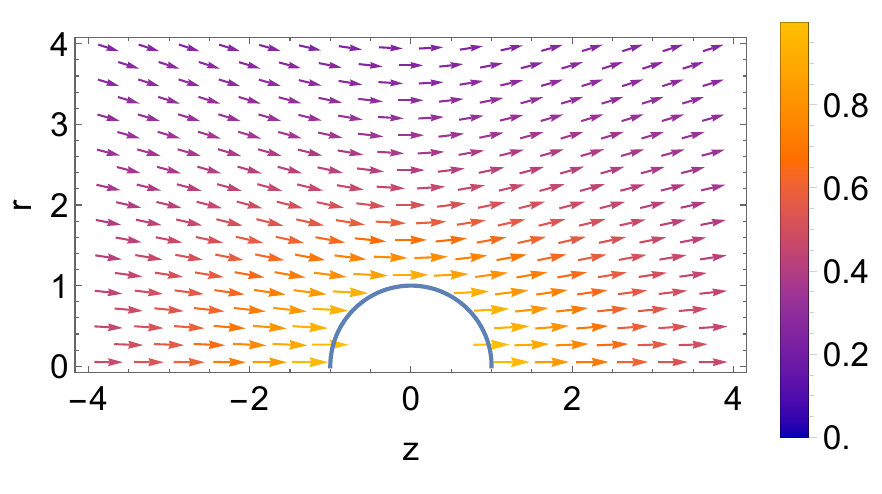}\caption{The scaled velocity field $\boldsymbol{v}_{\hat{h}=0}^{\left(1\right)}/U$
for $\hat{h}=0$ in the $z-x$ plane. The arrow color indicates the
magnitude of the local scaled velocity.}
\label{v for h=00003D00003D0} 
\end{figure}

We denote the position of center of the particle with $\boldsymbol{r}_{G}\left(t\right)=\left(0,0,Ut+z_{G,0}\right)$
and define the comoving frame as 
\begin{equation}
\boldsymbol{r}'=\boldsymbol{r}-\boldsymbol{r}_{G}\left(t\right).
\end{equation}
Hereafter in this section all the spatial coordinate and derivatives
are taken in this comoving frame. At the steady state viewed from
the comoving frame, the velocity and the composition fields are stationary:
\begin{equation}
0=\bar{\eta}\nabla^{2}\boldsymbol{v}-\psi\nabla\mu-\nabla p,\label{eq:stokes eq}
\end{equation}
\begin{equation}
\left(\boldsymbol{v}-\varepsilon U\boldsymbol{e}_{z}\right)\cdot\nabla\psi=\nabla\cdot\left[L\left(\psi\right)\nabla\mu\right].\label{eq:eqforpsi}
\end{equation}
Up to the first order of $\varepsilon$, we expand the velocity field, {}{composition field}
and chemical potential field as 
\begin{equation}
\boldsymbol{v}\left(\boldsymbol{r}\right)=\varepsilon\boldsymbol{v}^{\left(1\right)}\left(\boldsymbol{r}\right),
\end{equation}
\begin{equation}
{}{\psi=\psi^{\left(0\right)}+\varepsilon\psi^{\left(1\right)}},
\end{equation}
\begin{equation}
\mu=\mu^{\left(0\right)}+\varepsilon\mu^{\left(1\right)},
\end{equation}
{}{where $\psi^{\left(0\right)}$ is the equilibrium profile introduced below Eq. (\ref{eq:boundary_r_infty_psi}) and $\mu^{\left(0\right)}=0$ when the bulk composition is the critical one.} 
We introduce polar coordinate $(r,\theta,\varphi)$ with its origin
located at the center of the particle as 
\begin{equation}
\boldsymbol{r}'=\left(r\sin\theta\cos\varphi,r\sin\theta\sin\varphi,r\cos\theta\right).
\end{equation}
From the axisymmetry along the $z$ axis, in terms of a function $Q_{10}\left(r\right)$
that only depends on $r$, the lowest order of the chemical potential
deviation is given by 
\begin{equation}
\mu^{\left(1\right)}=Q_{10}\left(r\right)Y_{10}\left(\theta\right),
\end{equation}
where $Y_{10}\left(\theta\right)=\sqrt{3/\left(4\pi\right)}\cos\theta$
is the spherical harmonics of degree 1 and order 0. The velocity field
can be expressed as 
\begin{equation}
\boldsymbol{v}\left(\boldsymbol{r}'\right)=v_{\theta}\left(\boldsymbol{r}'\right)\boldsymbol{e}_{\theta}+v_{r}\left(\boldsymbol{r}'\right)\boldsymbol{e}_{r},\label{eq:velvrvt}
\end{equation}
where we write $\boldsymbol{e}_{r}$ and $\boldsymbol{e}_{\theta}$
for the unit vectors along the respective coordinate curves. From
the incompressibility condition, we can eliminate the pressure field
and $v_{r}\left(\boldsymbol{r}'\right)$ and $v_{\theta}\left(\boldsymbol{r}'\right)$
can be expressed as 
\begin{equation}
v_{r}\left(\boldsymbol{r}'\right)=R_{10}\left(r\right)Y_{10}\left(\theta\right),\:v_{\theta}\left(\boldsymbol{r}'\right)=\frac{1}{2r}\partial_{r}\left(r^{2}R_{10}\left(r\right)\right)\partial_{\theta}Y_{10}\left(\theta\right)\label{eq:vrvt}
\end{equation}
in terms of a function $R_{10}\left(r\right)$ that only depends on
$r$. We introduce dimensionless radial variable $\rho\equiv r/a$
and dimensionless functions 
\begin{equation}
\mathcal{Q}\left(\rho\right)=\frac{Q_{10}\left(r\right)\sqrt{L\left(0\right)}}{U\sqrt{\bar{\eta}}}\sqrt{\frac{3}{20\pi}},
\end{equation}
\begin{equation}
\mathcal{R}\left(\rho\right)=\frac{R_{10}\left(r\right)}{U}\sqrt{\frac{3}{4\pi}},
\end{equation}
\begin{equation}
\Psi\left(\rho\right)=-\frac{r^{2}}{3\sqrt{5\bar{\eta}L\left(0\right)}}\frac{d\psi^{\left(0\right)}\left(r\right)}{dr}.
\end{equation}
Substituting the definitions of $\mathcal{Q}\left(\rho\right)$ and
$\mathcal{R}\left(\rho\right)$ into Eqs. (\ref{eq:stokes eq}) and
(\ref{eq:eqforpsi}), we can obtain a set of ordinary differential
equations that $\mathcal{Q}\left(\rho\right)$ and $\mathcal{R}\left(\rho\right)$
satisfy 
\begin{equation}
\left(\rho\partial_{\rho}+1\right)\left(\rho\partial_{\rho}-2\right)\left(\rho\partial_{\rho}+3\right)\rho\partial_{\rho}\mathcal{R}\left(\rho\right)=-30\Psi\left(\rho\right)\mathcal{Q}\left(\rho\right),\label{eq:eqforR}
\end{equation}
\begin{equation}
\left(\rho\partial_{\rho}-1\right)\left(\rho\partial_{\rho}+2\right)\mathcal{Q}\left(\rho\right)=-3\Psi\left(\rho\right)\left[A\left(\rho\right)\left(\mathcal{R}\left(\rho\right)-1\right)-B\left(\rho\right)\partial_{\rho}\mathcal{Q}\left(\rho\right)\right].\label{eq:eqforQ}
\end{equation}
under the following boundary conditions at $\rho=1$, which are consistent
with Eqs. (\ref{eq:no-slip}) and {}{(\ref{eq:nmupr})}, and asymptotic
behaviors as $\rho\rightarrow\infty$ 
\begin{equation}
\mathcal{R}\left(1\right)=1,{}{\partial_{\rho}\mathcal{R}}\left(\rho=1\right)=0\;\mathrm{and\;\mathcal{R}\left(\rho\right)\rightarrow\mathcal{A}/\rho\;\mathrm{as\;\rho\rightarrow\infty}}\label{eq:bc for Rrho}
\end{equation}
\begin{equation}
{}{\partial_{\rho}\mathcal{Q}}\left(\rho=1\right)=0\,\mathrm{and\,}\mathcal{Q}\left(\rho\right)\rightarrow-\mathcal{B}/\rho^{2}\,\mathrm{as\,\rho\rightarrow\infty},\label{eq:bc for Qrho}
\end{equation}
where the amplitudes $\mathcal{A}$ and $\mathcal{B}$ should be determined
so that regular solutions $\mathcal{Q}\left(\rho\right)$ and $\mathcal{R}\left(\rho\right)$
defined for $1\leq\rho<\infty$ exist as discussed later. These asymptotic
behaviors as $\rho\rightarrow\infty$ are consistent with Eqs. (\ref{eq:eqforR})
and (\ref{eq:eqforQ}) since $\Psi\left(\rho\right)\propto\exp\left(-\rho a/\xi_{\infty}\right)$
as $\rho\rightarrow\infty$. Moreover, the asymptotic behavior $\mathcal{R}\left(\rho\right)\rightarrow\mathcal{A}/\rho$
as $\rho\rightarrow\infty$ is consistent with the fact that the velocity
field decays in proportion to $1/r$ when a net non-zero force is
acting on the particle. In Eqs. (\ref{eq:eqforR}) and (\ref{eq:eqforQ}),
the functions $A\left(\rho\right)$ and $B\left(\rho\right)$ are
defined as 
\begin{equation}
A\left(\rho\right)=\frac{L\left(0\right)}{L\left(\psi^{\left(0\right)}\left(a\rho\right)\right)}\;\mathrm{and\;}B\left(\rho\right)=\frac{L'\left(\psi^{\left(0\right)}\left(a\rho\right)\right)}{aL\left(\psi^{\left(0\right)}\left(a\rho\right)\right)}\sqrt{5\bar{\eta}L\left(0\right)}
\end{equation}
For $h=0$, $\Psi\left(\rho\right)$ vanishes and we can find the
solution $\mathcal{R}_{h=0}\left(\rho\right)$ for Eq. (\ref{eq:eqforR})
as 
\begin{equation}
\mathcal{R}_{h=0}\left(\rho\right)=\frac{3}{2\rho}-\frac{1}{2\rho^{3}}.\label{eq:R0}
\end{equation}
The scaled velocity field $\boldsymbol{v}\left(\boldsymbol{r}\right)/U$
for $h=0$, where the preferential adsorption vanishes, calculated
from Eqs. (\ref{eq:velvrvt}), (\ref{eq:vrvt}) and (\ref{eq:R0})
is plotted in Fig. \ref{v for h=00003D00003D0}. This scaled velocity
field, which leads to Stokes' law, was originally derived as the steady
flow field around a spherical particle moving at a constant velocity
in one-component fluid.

Using Lorenz's reciprocity of Stokesian hydrodynamics, the following
formula on the drag coefficient in terms of $\mathcal{Q}\left(\rho\right)$
is shown \citep{yabunaka2020drag,fujitani2018osmotic} 
\begin{equation}
\bar{\gamma}=6\pi\bar{\eta}a\left(1+\frac{10}{3}\int_{1}^{\infty}\alpha_{0}\left(\rho\right)\mathcal{Q}\left(\rho\right)\Psi\left(\rho\right)d\rho\right),\label{eq:formula}
\end{equation}
where the function $\alpha_{0}\left(\rho\right)$ is defined as 
\begin{equation}
\alpha_{0}\left(\rho\right)\equiv\frac{3}{2\rho}-\frac{1}{2\rho^{3}}-1=\mathcal{R}_{h=0}\left(\rho\right)-1.
\end{equation}
In \citep{okamoto2013drag}, in order to find the force acting on
the particle due to the order parameter deviation, they integrate
the surface stress tensor, whose explicit expression is not given
in this paper, on the particle surface. Moreover the surface stress
tensor depends on the deviation of the order parameter deviation profile
$\psi^{\left(1\right)}$, which is somewhat cumbersome to find from
the chemical deviation profile $\mu^{\left(1\right)}$ obtained by
solving Eqs. (\ref{eq:eqforR}) and (\ref{eq:eqforQ}). Therefore
Eq. (\ref{eq:formula}) is useful since we can skip this process of
integrating the surface stress tensor and calculate directly the drag
coefficient once we have the chemical deviation profile $\mu^{\left(1\right)}$
in terms of $\mathcal{Q}\left(\rho\right)$, {}{as is done in \citep{yabunaka2020drag,fujitani2018osmotic}}. 

\section{Numerical methods}

\subsection{Equilibrium Profiles}

We rescale the reduced temperature $\tau$ and the order parameter
field $\psi$ using characteristic values associated with the particle
radius $a$ 
\begin{equation}
\tau_{a}=\left(\frac{\xi_{0}}{a}\right)^{1/\nu},\quad\psi_{a}=\frac{\tau_{a}^{\beta}}{\sqrt{C_{2}}}.
\end{equation}
We then introduce the scaled variables 
\begin{align}
\hat{\tau} & =\frac{\tau}{\tau_{a}},\quad\hat{\psi}(\rho)=\frac{\psi(a\rho)}{\psi_{a}},\\
\hat{h} & =\frac{ha\sqrt{C_{2}}}{T_{c}C_{1}\tau_{a}^{\beta}}.
\end{align}
In typical experimental conditions, $\hat{h}$ is larger than 10~\citep{yabunaka2020drag}.
We also define the scaled distance from the critical point as $\hat{w}=w/\tau_{a}$.

Given $\hat{\tau}$ and $\hat{h}$, the equilibrium profile $\hat{\psi}^{(0)}(\rho)$
around a spherical particle satisfies
\begin{equation}
\left(\partial_{\rho}^{2}+\frac{2}{\rho}\partial_{\rho}\right)\hat{\psi}^{(0)}(\rho)=\frac{\left[2-\alpha+4(1-\alpha)\hat{\tau}\hat{w}^{-1}+5\alpha\hat{\tau}^{2}\hat{w}^{-2}\right]\hat{w}^{\gamma}}{6\left[2\beta+(1-2\beta)\hat{\tau}\hat{w}^{-1}\right]}\hat{\psi}^{(0)}(\rho),
\end{equation}
\begin{equation}
\hat{w}=\hat{\tau}+\hat{w}^{1-2\beta}\left(\hat{\psi}^{(0)}(\rho)\right)^{2}.
\end{equation}
The boundary conditions are 
\begin{align}
\hat{\psi}^{(0)}(\rho) & \rightarrow0\quad\text{as }\rho\to\infty,\\
\partial_{\rho}\hat{\psi}^{(0)}(\rho) & =-\hat{h}\hat{w}^{\nu\eta}\quad\text{at }\rho=1.
\end{align}

\begin{figure}
\centering \includegraphics[scale=0.8]{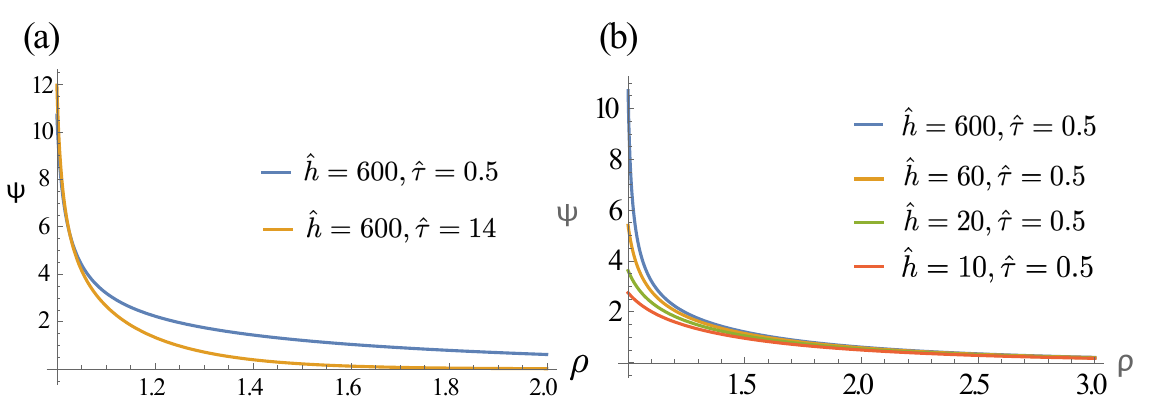} \caption{Equilibrium order parameter profiles $\hat{\psi}^{(0)}(\rho)$ for
selected values of $\hat{\tau}$ and $\hat{h}$. (a) Profiles showing
the exponential decay with different $\xi_{\infty}$. (b) Impact of
increasing $\hat{h}$ in the strong adsorption regime.}
\label{op_profile} 
\end{figure}

Examples of order parameter profiles for various $\hat{\tau}$ and
$\hat{h}$ have been calculated in~\citep{yabunaka2020drag}. Figure~\ref{op_profile}
shows the profiles for $(\hat{\tau},\hat{h})=(0.5,600),(14,600)$, {}{$(0.5,60)$, $(0.5,20)$
and $(0.5,10)$.} In the parameter range of interest, the strong adsorption
regime is realized 
\begin{equation}
\hat{h}^{2/3}\frac{\xi_{\infty}}{a}\gtrsim1.
\end{equation}
Near the particle surface at $\rho=1$, an adsorption layer of thickness
$\sim\xi_{\infty}/a$ forms due to preferential adsorption. Inside
this layer, the profile becomes off-critical with $\hat{\psi}^{(0)}(\rho)\gtrsim1$
and closely resembles the critical adsorption profile at $\hat{\tau}=0$.
Far from the surface ($\rho-1\gg\xi_{\infty}/a$), the order parameter
decays exponentially as 
\begin{equation}
\hat{\psi}^{(0)}(\rho)\propto\exp\left(-\frac{a\rho}{\xi_{\infty}}\right),
\end{equation}
as shown in Fig.~\ref{op_profile}(a). The scaled bulk correlation
lengths are $\xi_{\infty}/a=1.54$ and $0.191$ for $\hat{\tau}=0.5$
and $14$, respectively.

The analytical result at $\hat{\tau}=0$ based on local renormalized
functional theory~\citep{yabunaka2017critical} gives $\hat{\psi}^{(0)}(1)=3^{1/6}\hat{h}^{1/3}$
at $\eta=0$, which evaluates to 4.70 and 10.2 for $\hat{h}=60$ and
600, respectively. These are close to numerical values
of $\hat{\psi}^{(0)}(1)$ at $\hat{\tau}=0.5$.

As shown in Fig.~\ref{op_profile}(b), in the strong adsorption regime,
increasing $\hat{h}$ does not significantly affect the profile except
very close to the particle surface. In contrast, in the weak adsorption
regime $\hat{h}^{2/3}\xi_{\infty}/a\ll1$ treated in~\citep{okamoto2013drag},
the profile scales linearly with $\hat{h}$ for all $\rho\geq1$.

\subsection{Nonequilibrium profiles}

For each $\hat{\tau}$ (or $\hat{\xi}_{\infty}$), substituting the equilibrium profile $\hat{\psi}^{\left(0\right)}\left(\rho\right)$, we evaluate the profiles $A\left(\rho\right)$, $B\left(\rho\right)$, and $\Psi\left(\rho\right)$. These are needed to solve the profiles $\mathcal{Q}\left(\rho\right)$ and $\mathcal{R}\left(\rho\right)$ from Eqs.~(\ref{eq:eqforR}) and (\ref{eq:eqforQ}). The profiles $\mathcal{Q}\left(\rho\right)$ and $\mathcal{R}\left(\rho\right)$ for each $\hat{\tau}$ (or $\hat{\xi}_{\infty}$) were previously calculated by solving integral equations equivalent to Eqs.~(\ref{eq:eqforR}) and (\ref{eq:eqforQ}) \citep{fujitani2018osmotic,yabunaka2020drag}. However, iterative solutions of these integral equations converge only in a limited range of $\hat{\xi}_{\infty}$ \citep{yabunaka2020drag}, which is narrower than the range investigated in experiments \citep{omari2009effect}, as discussed in the next section.

\begin{figure}
    \includegraphics[scale=0.78]{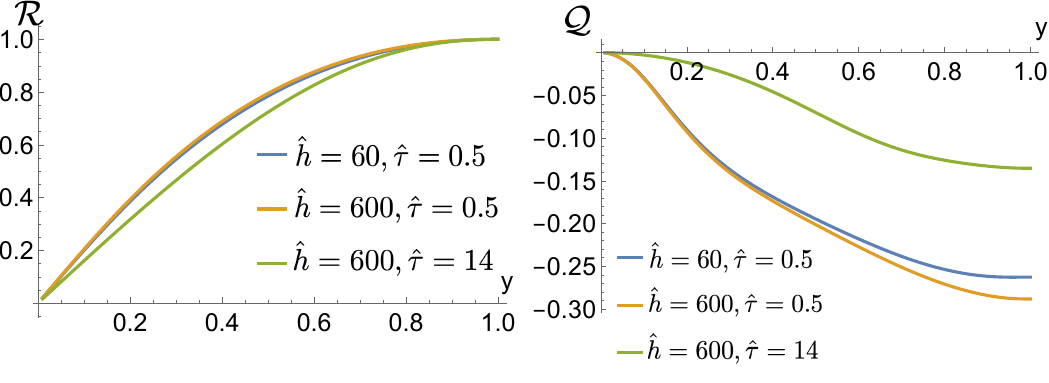} 
    \caption{Solutions $\tilde{\mathcal{Q}}\left(y\right)$ and $\tilde{\mathcal{R}}\left(y\right)$ for {}{$(\hat{h}, \hat{\tau})=(600, 0.5), (600, 14)$ and $(60, 0.5)$.} 
    These profiles are derived by solving the reformulated ODEs described in the text.}
    \label{sol Q and R}
\end{figure}

To overcome this difficulty, we directly solve Eqs.~(\ref{eq:eqforR}) and (\ref{eq:eqforQ}) as ODEs using the following compactified radial variable
\begin{equation}
y = \frac{1}{\rho}.
\end{equation}
A similar variable is used in \citep{fukuda2005friction}  to study the drag coefficient of a colloidal particle in the isotropic phase of a liquid crystal and we expect that it helps accurately resolve the hydrodynamic fields around the particle, which decays slowly (algebraically) as a function of $\rho$. In terms of $y$, the domain of interest becomes $\left(0, 1\right]$ instead of $\left[1, \infty\right)$ in terms of $\rho$. We define
\begin{align}
\tilde{\mathcal{R}}\left(y\right) &= \mathcal{R}\left(1/y\right), \\
\tilde{\mathcal{Q}}\left(y\right) &= \mathcal{Q}\left(1/y\right).
\end{align}
From Eqs.~(\ref{eq:eqforR}) and (\ref{eq:eqforQ}), the corresponding ODEs satisfied by $\tilde{\mathcal{Q}}\left(y\right)$ and $\tilde{\mathcal{R}}\left(y\right)$ are
\begin{equation}
y\tilde{\mathcal{Q}}''(y)-2\tilde{\mathcal{Q}}(y)+3\tilde \Psi(y)\left(\tilde{A}\left(y\right)(\tilde{\mathcal{R}}\left(y\right)-1)+y^{2}\tilde{B}\left(y\right)\tilde{\mathcal{Q}}'(y)\right)=0,\label{eq:eqforQ(y)}
\end{equation}
\begin{equation}
  y\left(y\frac{d^{4}\mathcal{\tilde{R}}}{dy^{4}}+4\frac{d^{3}\mathcal{\tilde{R}}}{dy^{3}}\right)-4\mathcal{\tilde{R}}''(y)+\frac{30\mathcal{\tilde{Q}}(y)\tilde \Psi(y)}{y^{2}}=0,\label{eq:eqforR(y)}
\end{equation}
where $\tilde{A}(y) = A(1/y)$, $\tilde{B}(y) = B(1/y)$ and  $\tilde \Psi(y)= \Psi(1/y)$. 

We seek solutions $\tilde{\mathcal{R}}\left(y\right)$ and $\tilde{\mathcal{Q}}\left(y\right)$ defined for $0 < y \leq 1$, satisfying the following boundary conditions
\begin{align}
\tilde{\mathcal{R}}(1) &= 1, \quad \tilde{\mathcal{R}}'(1) = 0, \\
\tilde{\mathcal{Q}}'(1) &= 0,
\end{align}
and the leading asymptotic behaviors
\begin{align}
\tilde{\mathcal{R}}\left(y\right) &\sim \mathcal{A} y, \label{eq:asympR} \\
\tilde{\mathcal{Q}}\left(y\right) &\sim -\mathcal{B} y^{2}, \label{eq:asympQ}
\end{align}
as $y \to 0$. These asymptotics suggest that $\tilde{\mathcal{R}}\left(y\right)$ and $\tilde{\mathcal{Q}}\left(y\right)$ can be smoothly extended to $y = 0$, allowing us to numerically determine the amplitudes $\mathcal{A}$ and $\mathcal{B}$. 

{}{For $(\hat{h}, \hat{\tau})=(600, 0.5), (600, 14)$ and $(60, 0.5)$, solutions $\tilde{\mathcal{Q}}\left(y\right)$ and $\tilde{\mathcal{R}}\left(y\right)$ were obtained with $(\mathcal{A}, \mathcal{B})=(2.07, 6.13), (1.61, 0.573)$ and $(2.02,6.00)$, respectively,} as shown in Fig.~\ref{sol Q and R}. See Appendix~\ref{sec:Numerics} for numerical details. {}{We can see that  $\tilde{\mathcal{R}}$ and  $\tilde{\mathcal{Q}}$ change only moderately when we increase $\hat{h}$ from 60 to 600, while $\psi(\rho=1)$ changes largely. This suggests that $\tilde{\mathcal{R}}$ and  $\tilde{\mathcal{Q}}$ are not very sensitive to the changes of $\psi$ very close to the solid surface $\rho=1$.}

The scaled drag coefficient is evaluated by substituting $\tilde{\Psi}(y)$ and $\tilde{\mathcal{Q}}(y)$ into
\begin{equation}
\bar{\gamma} = 6\pi\bar{\eta}a\left(1 + \frac{10}{3} \int_{0}^{1} \tilde{\alpha}_{0}(y)\tilde{\mathcal{Q}}(y)\tilde{\Psi}(y)\frac{dy}{y^{2}}\right), \label{eq:formuladrag}
\end{equation}
which is equivalent to Eq.~(\ref{eq:formula}).

\section{Results}

\begin{figure}
\includegraphics[scale=0.5]{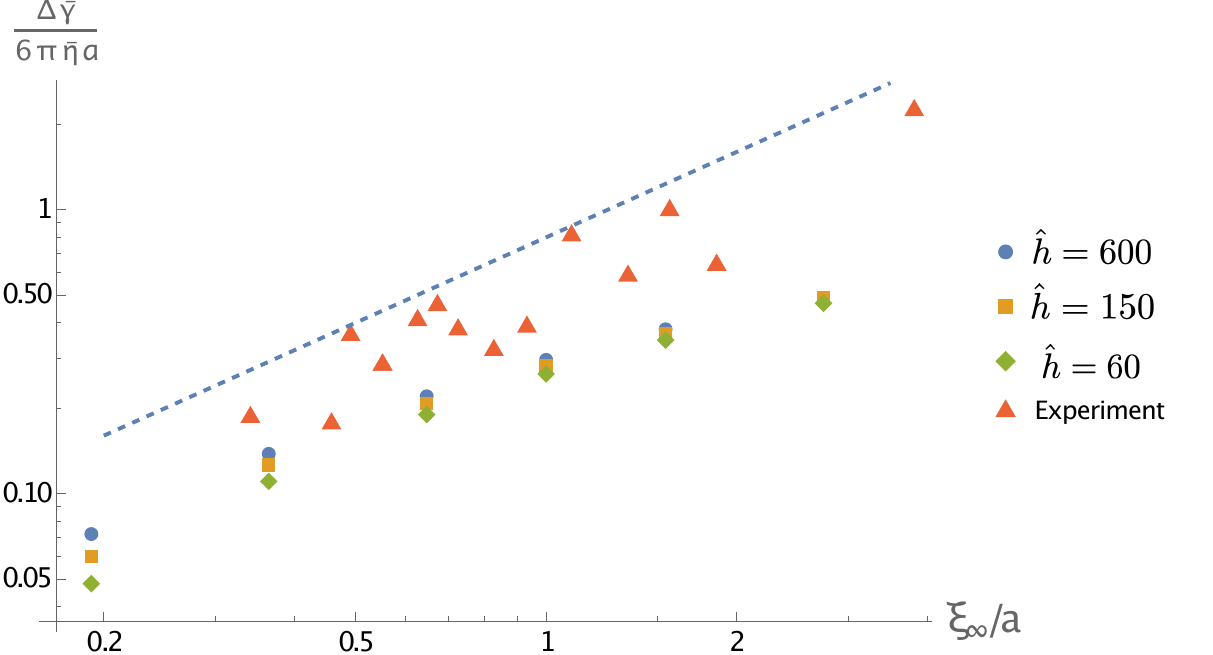} \caption{{}{A log-log plot of} scaled deviation of the drag coefficient, $\Delta\bar{\gamma}/(6\pi a\bar{\eta})$,
plotted against $\xi_{\infty}/a$ for $\hat{h}=600$, 150, and 60.
The experimental estimates {}{with the particle radius $a=25 \rm{nm}$} from Ref.~\citep{omari2009effect} are
also shown. The dashed line represents $\Delta\bar{\gamma}/(6\pi a\bar{\eta})=\xi/a$
as a visual guide.}
\label{dragfig} 
\end{figure}

In Fig.~\ref{dragfig}, we plot the deviation of the drag coefficient
from Stokes' law, $\Delta\bar{\gamma}\equiv\bar{\gamma}-6\pi\bar{\eta}a$,
as a function of the scaled bulk correlation length $\xi_{\infty}/a=\hat{\tau}^{-\nu}$,
and compare it with the experimental estimates {}{for the particle radius $a=25 \rm{nm}$} reported by Omari et
al.~\citep{omari2009effect}{}{, where the diffusion constant of a colloidal particle $D$ is measured by means of light scattering and the drag coefficient is estimated via the Stokes-Eistein relation $D=k_{B}T/\bar{\gamma}$. $\xi_\infty$ in their system is estimated as 0.25 $\tau^{-\nu} \, {\rm nm}$, where  $\tau$ is written as $t$ in their notation.} {}{We note that the final experimental data point, $(\xi_{\infty}/a, \,\Delta\bar{\gamma}/6\pi\bar{\eta}a) = (3.82, \, 2.30)$, has a significantly larger uncertainty (approximately $40\%$) for $\Delta\bar{\gamma}/6\pi\bar{\eta}a$, compared to other data points, which have uncertainties of at most $10\%$ for $\Delta\bar{\gamma}/6\pi\bar{\eta}a$. Therefore the linear dependence of $\Delta\bar{\gamma}/6\pi\bar{\eta}a$ as a function of $\xi_{\infty}/a$ is more robust typically for $0.3\lesssim\xi_{\infty}/a\lesssim2$}.  The quantity $\Delta\bar{\gamma}$
increases with both $\hat{h}$ and $\xi_{\infty}/a$, which is reasonable
since the deviation originates from the adsorption layer that forms
when $\hat{h}\neq0$ and grows thicker for larger $\xi_{\infty}/a$.

In \citep{yabunaka2020drag}, $\Delta\bar{\gamma}$ was calculated
only up to $\xi_{\infty}/a<0.6$ and $\xi_{\infty}/a<0.4$ for $\hat{h}=60$
and $150$, respectively. Our present numerical scheme, which directly
solves for $\tilde{\mathcal{R}}(y)$ and $\tilde{\mathcal{Q}}(y)$,
allows us to compute $\Delta\bar{\gamma}$ over a broader range of
$\xi_{\infty}/a$. We have verified that our method reproduces the
results from \citep{yabunaka2020drag} in their respective regimes.

The experimental values exhibit some scatter but are generally larger
than the theoretical predictions. In the intermediate regime $0.2\lesssim\xi_{\infty}/a\lesssim1$,
the dependence of $\Delta\bar{\gamma}$ on $\xi_{\infty}/a$ appears
nearly linear. However, for larger values $\xi_{\infty}/a\gtrsim1$,
the increase in $\Delta\bar{\gamma}$ becomes more gradual. {}{The fluctuation-dissipation relation indicates that accounting for all fluctuations in the system, including the Brownian motion of the particle, is crucial for accurately determining the dissipation. However, the current framework, which is based on local renormalized functional theory, may overlook some of these contributions (see the next section) and underestimate $\bar{\gamma}$, potentially explaining the observed discrepancy.}

The dependence of $\Delta\bar{\gamma}$ on $\hat{h}$ is relatively
weak for $\hat{h}\geq60$ within the range $0.2\lesssim\xi_{\infty}/a\lesssim2$,
and diminishes further for larger $\xi_{\infty}/a$. As noted earlier,
the equilibrium composition profile remains in the strong adsorption
regime for these parameter ranges. Therefore, increasing $\hat{h}$ {}{in the parameter range considered here}
does not almost affect the profile except extremely close to
the particle surface {}{and not significantly affect $\mathcal{R}$ and $\mathcal{Q}$ for $0\le y \le 1$ as shown in Fig. \ref{sol Q and R}.} This suggests that $\Delta\bar{\gamma}$ is
not very sensitive to changes in the profile very close to the surface.
However, further data at even larger $\hat{h}$, which are numerically
demanding to obtain, may be necessary to clarify this dependence.

\begin{figure}
\includegraphics[scale=0.5]{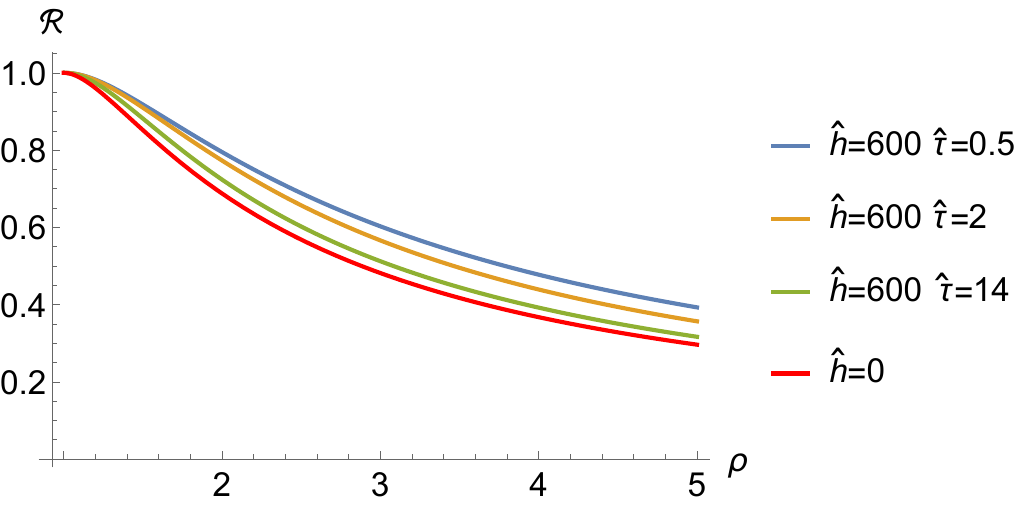} \caption{Radial component $\mathcal{R}(\rho)$ of the scaled velocity for $\hat{\tau}=0.5$,
2, and 14 at $\hat{h}=600$.}
\label{Rrho} 
\end{figure}

In Fig.~\ref{Rrho}, we show the function $\mathcal{R}(\rho)$, which
corresponds to the radial component of the scaled velocity field {[}see
Eq.~(\ref{eq:vrvt}){]}. The function $\mathcal{R}(\rho)$ decays
smoothly as $\rho$ increases, contradicting a rigid-layer interpretation
of the adsorption layer. For smaller values of $\hat{\tau}$, where
the adsorption layer becomes thicker, $\mathcal{R}(\rho)$ exhibits
larger values and decays more slowly.

\begin{figure}
\includegraphics[scale=0.5]{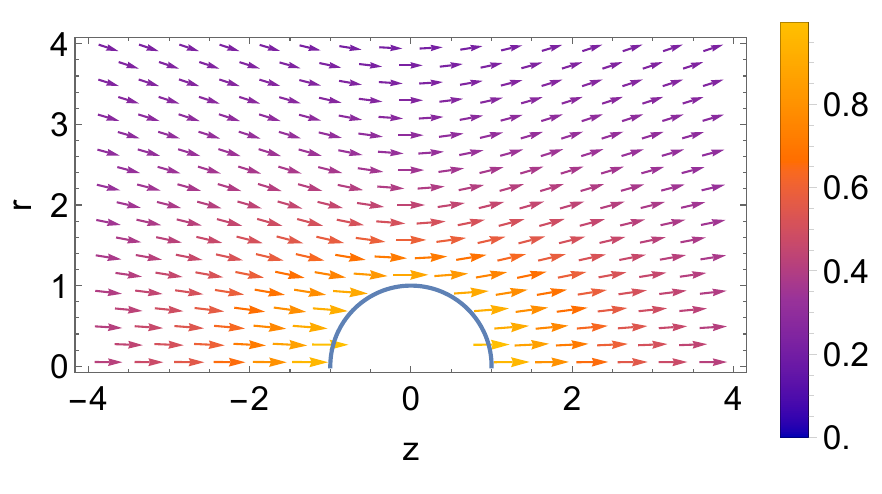} \includegraphics[scale=0.5]{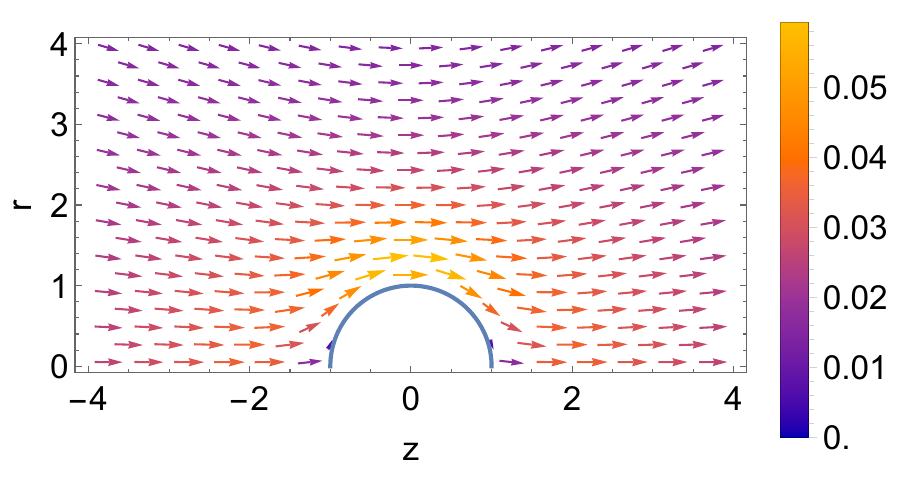}
\caption{(Left) Scaled velocity field $\boldsymbol{v}_{\hat{h}=600,\hat{\tau}=14}^{(1)}/U$
in the $z$--$x$ plane. (Right) Its deviation from the no-adsorption
field $\boldsymbol{v}_{\hat{h}=0}^{(1)}/U$. The arrow color indicates
the magnitude of the local scaled velocity.}
\label{velocity field} 
\end{figure}

\begin{figure}
\includegraphics[scale=0.5]{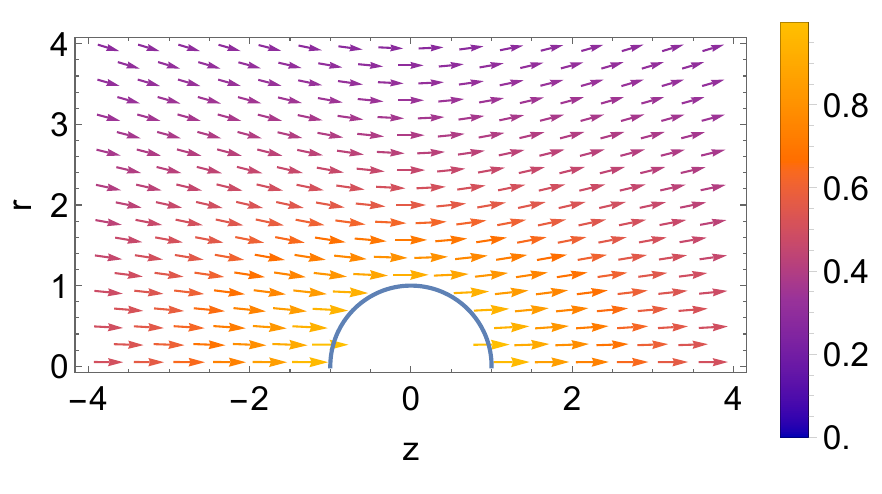} \includegraphics[scale=0.5]{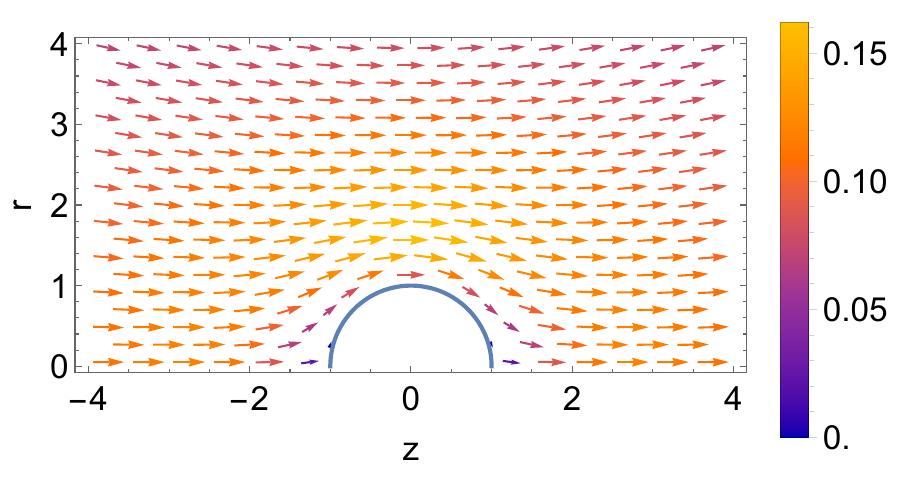}
\caption{(Left) Scaled velocity field $\boldsymbol{v}_{\hat{h}=600,\hat{\tau}=0.5}^{(1)}/U$
in the $z$--$x$ plane. (Right) Its deviation from the no-adsorption
field $\boldsymbol{v}_{\hat{h}=0}^{(1)}/U$. The arrow color indicates
the magnitude of the local scaled velocity.}
\label{velocity field2} 
\end{figure}

\begin{figure}
\includegraphics[scale=0.5]{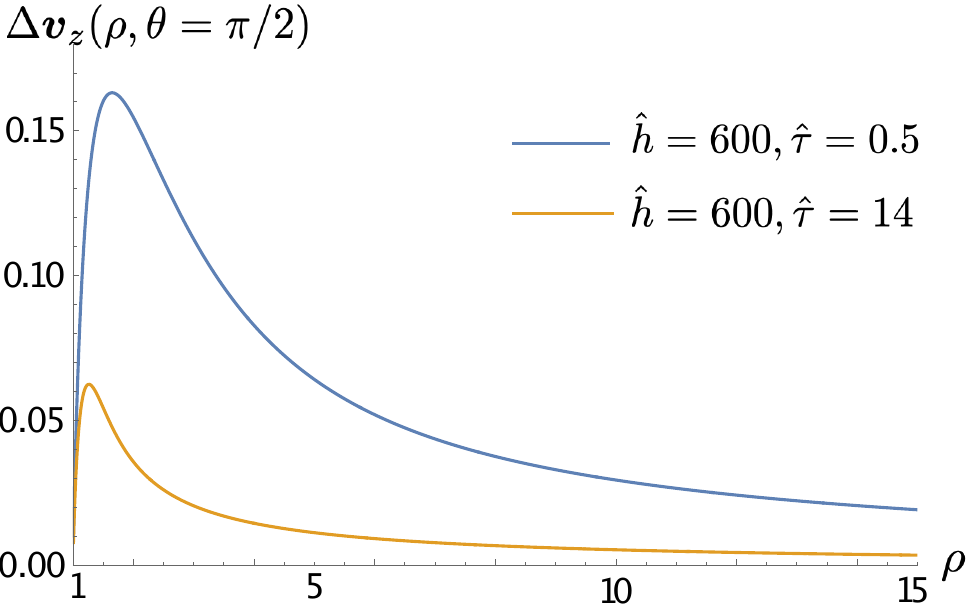}\caption{{}{The $z$-component of the deviation in the scaled velocity field $\boldsymbol{v}_{\hat{h}=600, \hat{\tau}=0.5,14}^{(1)}$ compared to the no-adsorption field $\boldsymbol{v}_{\hat{h}=0}^{(1)}/U$.}}
\label{dvz} 
\end{figure}

Figs.~\ref{velocity field} and \ref{velocity field2} present the
scaled velocity field $\boldsymbol{v}^{(1)}/U$ around the particle
for $\hat{\tau}=14$ and $0.5$ with $\hat{h}=600$, along with their
deviations from the no-adsorption case, $(\boldsymbol{v}^{(1)}-\boldsymbol{v}_{h=0}^{(1)})/U$.
As $\hat{\tau}$ decreases, the magnitude $\left|\boldsymbol{v}^{(1)}-\boldsymbol{v}_{h=0}^{(1)}\right|/U$
increases and decays more slowly with $\rho$, indicating that the
particle entrains the surrounding fluid more strongly. {}{This can be clearly seen in Fig. \ref{dvz}, where the $z$-component of the deviation $(\boldsymbol{v}^{(1)}-\boldsymbol{v}_{h=0}^{(1)})/U$ at $\theta=\pi/2$ is plotted as a function of $\rho$.} 

\section{Conclusion and discussion}

We have successfully calculated the drag coefficient of a spherical
colloidal particle in a near-critical binary fluid mixture over a
wider range of the bulk correlation length $\xi_{\infty}$ than previously
reported, by employing local renormalized functional theory combined
with improved numerical techniques. While the values of the drag coefficient
$\bar{\gamma}$ obtained from our calculations remain somewhat smaller
than those inferred from experimental measurements \citep{omari2009effect},
the linear dependence of $\bar{\gamma}$ on the dimensionless correlation
length $\xi_{\infty}/a$, observed experimentally in the range $0.3\lesssim\xi_{\infty}/a\lesssim2$,
is largely reproduced. In particular, our results show that $\bar{\gamma}$
increases approximately linearly with $\xi_{\infty}/a$ for $0.2\lesssim\xi_{\infty}/a\lesssim1$,
while for $\xi_{\infty}/a\gtrsim1$ the increase becomes slightly
more gradual.

Local renormalized functional theory correctly captures the scaling
behavior associated with critical adsorption by incorporating renormalization
effects locally within the free energy functional. However, it does
not fully capture the impact of spatially distributed thermal noise
in the surrounding fluid leading to Brownian motion of the particle, which can influence the adsorption profile,
especially when $\xi_{\infty}/a\gtrsim1$ \citep{yabunaka2020drag}.
This omission may partly explain the remaining discrepancy between
our theoretical predictions and experimental results in that regime.
To accurately account for such effects, more systematic theoretical
frameworks, such as fluctuating hydrodynamics or molecular dynamics
simulations, may be required. {}{Moreover, we have assumed that the viscosity $\bar{\eta}$ is uniform, which can be justified when the composition remains very close to the critical value throughout the system. In more general cases, the viscosity may be nonuniform, and if the viscosity within the adsorption layer is lower than the bulk viscosity, lubrication effects could potentially reduce the drag coefficient $\bar{\gamma}$.} 

{}{The experimental estimates are based on the linear Langevin equation for the position of the center of the colloidal particle. The validity of such description in the presence of deformed adsorption layer, whose relaxation can be slow near the critical point,  should  be checked with more systematic theoretical
frameworks as well.}


We conclude with several remarks regarding the methodology and future
directions: (i) Our approach to solving the hydrodynamic equations
by reformulating them as ordinary differential equations (ODEs) using
a compactified radial variable proved more efficient and stable than
the recursive numerical schemes based on integral equations used in
previous studies \citep{fujitani2018osmotic, yabunaka2020drag}. Additionally,
the use of the concise expression of the drag coefficient derived
from Lorentz reciprocity \citep{yabunaka2020drag, fujitani2018osmotic}
allowed us to avoid directly integrating the stress tensor over the
particle surface, as was directly performed with the Gaussian free
energy \citep{okamoto2013drag}. (ii) The numerical techniques developed
here could be extended to other linear hydrodynamic problems involving
colloidal particles in complex fluids, such as those encountered in
active microrheology or phoretic phenomena (e.g., electrophoresis
and thermophoresis). Previous studies on these topics often use simplified
models of soft particles, introducing assumptions about the shape
and frictional properties of soft layers \citep{OHSHIMA1995189, lee2021diffusiophoresis, huang2012diffusiophoresis}.
Extension of our method may offer a more systematic and flexible approach that could
account for more detailed structural and hydrodynamic interactions.
For instance, diffusiophoresis of colloidal particles in near-critical
mixtures under composition gradients has already been modeled using
hydrodynamic equations similar to those employed in this study \citep{fujitani2022diffusiophoresis}.
(iii) Near the critical point, selective solvation of impurities such
as ions is known to significantly alter static properties of adsorption
and wetting layers \cite{okamoto2011charged,samin2011attraction}. It would be a natural extension of this work to
investigate how such effects modify the drag coefficient, particularly
in systems where ionic solvation couples strongly with compositional
fluctuations. {}{(iv) It would be interesting to extend this work to the cases with off-critical bulk compositions, where the critical casimir effects are known to be enhanced  either with \cite{okamoto2011charged,samin2011attraction} or without ions \citep{okamoto2012casimir} or a wetting layer can appear around the particle, depending on the mixture temperature.} 
\begin{acknowledgments}
This work is supported by JSPS Grant-in-Aid for Scientific Research
(Grants Nos. 21K03488 and 24K06984). The author is grateful to Profs.
Y. Fujitani and A. Onuki for illuminating discussion. 

\end{acknowledgments}

\bibliographystyle{plain}
\bibliographystyle{unsrt}
\bibliography{bib}

\begin{thebibliography}{10}

\bibitem{stokes1851effect}
George~Gabriel Stokes et~al.
\newblock On the effect of the internal friction of fluids on the motion of
  pendulums.
\newblock 1851.

\bibitem{einstein1905motion}
Albert Einstein.
\newblock On the motion of small particles suspended in a stationary liquid.
\newblock {\em Ann. Phys}, 322(8):549--560, 1905.

\bibitem{zia2018active}
Roseanna~N Zia.
\newblock Active and passive microrheology: Theory and simulation.
\newblock {\em Annual Review of Fluid Mechanics}, 50(1):371--405, 2018.

\bibitem{fukuda2005friction}
Jun-ichi Fukuda, Holger Stark, and Hiroshi Yokoyama.
\newblock Friction drag of a spherical particle in a liquid crystal above the
  isotropic-nematic transition.
\newblock {\em Physical Review E---Statistical, Nonlinear, and Soft Matter
  Physics}, 72(2):021701, 2005.

\bibitem{furukawa2004microrheology}
Akira Furukawa.
\newblock Microrheology of entangled polymer solutions.
\newblock {\em The Journal of chemical physics}, 121(19):9716--9732, 2004.

\bibitem{beysens1985adsorption}
D~Beysens and D~Esteve.
\newblock Adsorption phenomena at the surface of silica spheres in a binary
  liquid mixture.
\newblock {\em Physical review letters}, 54(19):2123, 1985.

\bibitem{furukawa2013nonequilibrium}
Akira Furukawa, Andrea Gambassi, Siegfried Dietrich, and Hajime Tanaka.
\newblock Nonequilibrium critical casimir effect in binary fluids.
\newblock {\em Physical review letters}, 111(5):055701, 2013.

\bibitem{okamoto2013attractive}
Ryuichi Okamoto and Akira Onuki.
\newblock Attractive interaction and bridging transition between neutral
  colloidal particles due to preferential adsorption in a near-critical binary
  mixture.
\newblock {\em Physical Review E---Statistical, Nonlinear, and Soft Matter
  Physics}, 88(2):022309, 2013.

\bibitem{gallagher1992partitioning}
PD~Gallagher and JV~Maher.
\newblock Partitioning of polystyrene latex spheres in immiscible critical
  liquid mixtures.
\newblock {\em Physical Review A}, 46(4):2012, 1992.

\bibitem{PhysRevLett.100.188303}
Hua Guo, Theyencheri Narayanan, Michael Sztuchi, Peter Schall, and Gerard~H.
  Wegdam.
\newblock Reversible phase transition of colloids in a binary liquid solvent.
\newblock {\em Phys. Rev. Lett.}, 100:188303, May 2008.

\bibitem{bonn2009direct}
Daniel Bonn, Jakub Otwinowski, Stefano Sacanna, Hua Guo, Gerard Wegdam, and
  Peter Schall.
\newblock Direct observation of colloidal aggregation by critical casimir
  forces.
\newblock {\em Physical review letters}, 103(15):156101, 2009.

\bibitem{wolynes1976osmotic}
Peter~G Wolynes.
\newblock Osmotic effects near the critical point.
\newblock {\em The Journal of Physical Chemistry}, 80(14):1570--1572, 1976.

\bibitem{PhysRevE.109.064610}
Shunsuke Yabunaka and Youhei Fujitani.
\newblock Thermo-osmosis of a near-critical binary fluid mixture: A general
  formulation and universal flow direction.
\newblock {\em Phys. Rev. E}, 109:064610, Jun 2024.

\bibitem{samin2017interplay}
Sela Samin and Ren{\'e} Van~Roij.
\newblock Interplay between adsorption and hydrodynamics in nanochannels:
  towards tunable membranes.
\newblock {\em Physical Review Letters}, 118(1):014502, 2017.

\bibitem{omari2009effect}
Rami~A Omari, Christopher~A Grabowski, and Ashis Mukhopadhyay.
\newblock Effect of surface curvature on critical adsorption.
\newblock {\em Physical review letters}, 103(22):225705, 2009.

\bibitem{floter1995universal}
G~Fl{\"o}ter and S~Dietrich.
\newblock Universal amplitudes and profiles for critical adsorption.
\newblock {\em Zeitschrift f{\"u}r Physik B Condensed Matter}, 97:213--232,
  1995.

\bibitem{hanke1999critical}
Andreas Hanke and S~Dietrich.
\newblock Critical adsorption on curved objects.
\newblock {\em Physical Review E}, 59(5):5081, 1999.

\bibitem{rudnick1982order}
Joseph Rudnick and David Jasnow.
\newblock Order-parameter profile in semi-infinite systems at criticality.
\newblock {\em Physical Review Letters}, 48(16):1059, 1982.

\bibitem{fisher1978wall}
Michael~E Fisher and Pierre-Gilles de~Gennes.
\newblock Wall phenomena in a critical binary mixture.
\newblock {\em CR Acad. Sci. Paris B}, 287(8):207--209, 1978.

\bibitem{okamoto2012casimir}
Ryuichi Okamoto and Akira Onuki.
\newblock Casimir amplitudes and capillary condensation of near-critical fluids
  between parallel plates: Renormalized local functional theory.
\newblock {\em The Journal of Chemical Physics}, 136(11), 2012.

\bibitem{FUJITANI2024114050}
Youhei Fujitani.
\newblock Preferential adsorption in a near-critical binary fluid mixture as
  analyzed in the framework of the non-random two-liquid model.
\newblock {\em Fluid Phase Equilibria}, 580:114050, 2024.

\bibitem{okamoto2013drag}
Ryuichi Okamoto, Youhei Fujitani, and Shigeyuki Komura.
\newblock Drag coefficient of a rigid spherical particle in a near-critical
  binary fluid mixture.
\newblock {\em journal of the physical society of japan}, 82(8):084003, 2013.

\bibitem{yabunaka2020drag}
Shunsuke Yabunaka and Youhei Fujitani.
\newblock Drag coefficient of a rigid spherical particle in a near-critical
  binary fluid mixture, beyond the regime of the gaussian model.
\newblock {\em Journal of Fluid Mechanics}, 886:A2, 2020.

\bibitem{onuki2002phase}
Akira Onuki.
\newblock {\em Phase transition dynamics}.
\newblock Cambridge University Press, 2002.

\bibitem{puri1997surface}
Sanjay Puri and Harry~L Frisch.
\newblock Surface-directed spinodal decomposition: modelling and numerical
  simulations.
\newblock {\em Journal of Physics: Condensed Matter}, 9(10):2109, 1997.

\bibitem{KAWASAKI19701}
Kyozi Kawasaki.
\newblock Kinetic equations and time correlation functions of critical
  fluctuations.
\newblock {\em Annals of Physics}, 61(1):1--56, 1970.

\bibitem{fujitani2018osmotic}
Youhei Fujitani.
\newblock Osmotic effects on dynamics of a colloidal rigid sphere in a
  near-critical binary fluid mixture.
\newblock {\em Journal of the Physical Society of Japan}, 87(8):084602, 2018.

\bibitem{yabunaka2017critical}
Shunsuke Yabunaka and Akira Onuki.
\newblock Critical adsorption profiles around a sphere and a cylinder in a
  fluid at criticality: Local functional theory.
\newblock {\em Physical Review E}, 96(3):032127, 2017.

\bibitem{OHSHIMA1995189}
Hiroyuki Ohshima.
\newblock Electrophoresis of soft particles.
\newblock {\em Advances in Colloid and Interface Science}, 62(2):189--235,
  1995.

\bibitem{lee2021diffusiophoresis}
Yu-Fan Lee, Wen-Chun Chang, Yvonne Wu, Leia Fan, and Eric Lee.
\newblock Diffusiophoresis of a highly charged soft particle in electrolyte
  solutions.
\newblock {\em Langmuir}, 37(4):1480--1492, 2021.

\bibitem{huang2012diffusiophoresis}
Ping~Y Huang and Huan~J Keh.
\newblock Diffusiophoresis of a spherical soft particle in electrolyte
  gradients.
\newblock {\em The Journal of Physical Chemistry B}, 116(25):7575--7589, 2012.

\bibitem{fujitani2022diffusiophoresis}
Youhei Fujitani.
\newblock Diffusiophoresis in a near-critical binary fluid mixture.
\newblock {\em Physics of Fluids}, 34(4), 2022.

\bibitem{okamoto2011charged}
Ryuichi Okamoto and Akira Onuki.
\newblock Charged colloids in an aqueous mixture with a salt.
\newblock {\em Physical Review E---Statistical, Nonlinear, and Soft Matter
  Physics}, 84(5):051401, 2011.

\bibitem{samin2011attraction}
Sela Samin and Yoav Tsori.
\newblock Attraction between like-charge surfaces in polar mixtures.
\newblock {\em Europhysics Letters}, 95(3):36002, 2011.

\end{thebibliography}

\appendix

\section{Details of Numerical Implementation}\label{sec:Numerics}

We first consider the simpler case without preferential adsorption
($h=0$), where the composition remains at its critical value $\psi=0$
throughout the domain $C^{e}$, implying $\mathcal{Q}(\rho)=0$. In
this case, Eq.~(\ref{eq:eqforR(y)}), a fourth-order ordinary differential
equation (ODE) for $\tilde{\mathcal{R}}(y)$, can be solved analytically
under the following boundary conditions 
\begin{equation}
\tilde{\mathcal{R}}(1)=1,\quad\tilde{\mathcal{R}}'(1)=0,\quad\tilde{\mathcal{R}}''(1)=R_{1},\quad\tilde{\mathcal{R}}'''(1)=R_{2}.
\end{equation}
The general solution under these conditions is given by 
\begin{equation}
\tilde{\mathcal{R}}(y)=\frac{R_{1}(20y^{5}-60y^{3}+40y^{2})+R_{2}(4y^{5}-10y^{3}+5y^{2}+1)+120y^{2}}{120y^{2}}.
\end{equation}
By choosing $R_{1}=-3$ and $R_{2}=0$, we obtain a smooth solution
over the interval $0\leq y\leq1$ 
\begin{equation}
\tilde{\mathcal{R}}(y)=\frac{3}{2}y-\frac{1}{2}y^{3},
\end{equation}
which matches Eq.~(\ref{eq:R0}). This implies that the asymptotic
amplitude $\mathcal{A}$ can be fixed at $3/2$ by requiring smooth,
analytic behavior as $y\to0$ (consistent with Eq.~(\ref{eq:asympR})).
For other values of $\mathcal{A}$, the solution diverges near $y=0$.

We now generalize the above procedure to the case of preferential
adsorption ($h\neq0$), where $\tilde{\mathcal{Q}}(y)$ and $\tilde{\mathcal{R}}(y)$
must be determined numerically from Eqs.~(\ref{eq:eqforQ(y)}) and~(\ref{eq:eqforR(y)}).
Here, Eq.~(\ref{eq:eqforQ(y)}) is a second-order ODE for $\tilde{\mathcal{Q}}(y)$.
To numerically implement the asymptotic behaviors described in Eqs.~(\ref{eq:asympR}) 
and~(\ref{eq:asympQ}) as $y\to0$, we solve the equations over the
domain $y\in(0,1]$, imposing the following boundary conditions:

For $\tilde{\mathcal{R}}(y)$ 
\begin{equation}
\tilde{\mathcal{R}}(1)=1,\quad\tilde{\mathcal{R}}'(1)=0,\quad\tilde{\mathcal{R}}(y_{\mathrm{ref}})=\mathcal{A}y_{\mathrm{ref}},\quad\tilde{\mathcal{R}}'(y_{\mathrm{ref}})=\mathcal{A},
\end{equation}
and for $\tilde{\mathcal{Q}}(y)$
\begin{equation}
\tilde{\mathcal{Q}}'(1)=0,\quad\tilde{\mathcal{Q}}(y_{\mathrm{ref}})=-\mathcal{B}y_{\mathrm{ref}}^{2},
\end{equation}
where $y_{\mathrm{ref}}$ is chosen to be sufficiently small (typically
$y_{\mathrm{ref}}\sim0.001$).

We solve these ODEs using Mathematica (version 14, Wolfram Research)
with the \texttt{NDSolve} function, applying the following numerical
options: 
\begin{quote}
\texttt{MaxStepSize}~$\rightarrow$~0.0001, \texttt{AccuracyGoal}~$\rightarrow$~15,
\texttt{WorkingPrecision}~$\rightarrow$~20, \texttt{InterpolationOrder}~$\rightarrow$~All,
\texttt{MaxSteps}~$\rightarrow~\infty$. 
\end{quote}
To determine the coefficients $\mathcal{A}$ and $\mathcal{B}$, we
employ a dichotomy (bisection) method, requiring that both $\tilde{\mathcal{R}}(y)$
and $\tilde{\mathcal{Q}}(y)$ exhibit smooth behavior near $y=0$,
consistent with the asymptotic forms in Eqs.~(\ref{eq:asympR}) and~(\ref{eq:asympQ}).
Although choosing a finite $y_{\mathrm{ref}}>0$ introduces slight
errors in the estimation of $\mathcal{A}$ and $\mathcal{B}$, we
estimate that, for $y_{\mathrm{ref}}\sim0.001$, the impact on the
scaled drag coefficient is smaller than $0.01\Delta\bar{\gamma}/(6\pi\bar{\eta}a)$. 
\end{document}